\documentclass[11pt]{article}  

\usepackage{amsmath,amsthm,latexsym,amssymb,amsfonts,epsfig}

\newcommand{\be}{\begin{equation}}
\newcommand{\ee}{\end{equation}}
\newcommand{\ba}{\begin{eqnarray}}
\newcommand{\ea}{\end{eqnarray}}

\title{{\sf Non-perturbative, background independent Fock representations for
canonical quantum gravity}}
\author{
{\sf T. Thiemann}$^1$\thanks{{\sf 
thomas.thiemann@gravity.fau.de}}\\
\\
{\sf $^1$ Inst. for Quantum Gravity, FAU Erlangen -- N\"urnberg,}\\
{\sf Staudtstr. 7, 91058 Erlangen, Germany}\\
}
\date{{\small\sf \today}}

\makeatletter
\@addtoreset{equation}{section}
\makeatother

\begin{document} 

\maketitle

{\sf

\begin{abstract}
A UV complete quantum field theory of general relativity is believed to require a 
non-perturbative approach. Moreover, background independence of classical general 
relativity supplies a physical selection for suitable Hilbert space representations
of the quantum geometry and matter fields.

In this contribution we show that, contrary to common intuition, there exist rigorous, background 
independent Fock representations available for a non-perturbative canonical quantisation of 
geometry and suitable matter fields. 

This is interesting because the Fock Hilbert space 
is separable while the Hilbert space of other manifestly background independent and 
non-perturbative canonical quantisation programmes is not. Since non-separability
is a source for quantisation ambiguities, such a Fock representation may help 
to arrive at a significantly more predictive theory. As a simple application
we discuss the cosmological truncation and mechanisms for quantum bounces.   

To make this manuscript concise we focus on the simplest incarnation of this idea. More 
details and many extensions are supplied in a companion paper. 
\end{abstract}

\section{Introduction}
\label{s1}

In the canonical approach to quantum gravity one considers foliations of the spacetime 
manifold by mutually diffeomorphic hypersurfaces and defines a phase space based on 
fields on those hypersurfaces. Due to general covariance, the corresponding 
so called primary Hamiltonian is a linear combination of first class constraints on 
that phase space. These generate gauge transformations on the fields. 
Observables are gauge invariant functions on the phase space. The dynamics of those 
observables is not generated by the primary Hamiltonian (which leaves them invariant 
by definition) but rather by a physical Hamiltonian. That physical Hamiltonian 
depends on the choice of a reference frame (see e.g. \cite{1} and references therein with 
a focus on applications in general relativity). The canonical quantisation 
of this classical field theory can now take two routes: In the first route, called
constraint quantisation, one quantises the entire phase space and the constraints 
as operators on a so called kinematical Hilbert space
and solves the quantum constraints by constructing the joint kernel of the quantum constraints.
The kernel must be equipped with a Hilbert space structure defining the physical 
Hilbert space which is supposed to carry a $\ast-$ representation of the algebra of observables.
In the second route, called reduced phase space quantisation, one solves the constraints 
classically and quantises only the algebra of observables directly on the physical Hilbert space.  

The motivation for the present work is the following situation in Loop Quantum Gravity (LQG) 
(see \cite{2} and references therein for all details) which is the leading candidate 
for canonical quantisation approach to general relativity in 3+1 spacetime dimensions. 
We describe first the constraint quantisation route: 
\begin{itemize}
\item[1.] {\it Choice of polarisation}\\
The classical phase space has many canonical charts. In LQG one chooses a canonical pair 
consisting of an su(2) valued connection $A$ and a su(2) valued vector density 
of weight one $E$ (densitised triad) on the spatial hypersurface as conjugate momentum in the geometry sector. 
Similarly, e.g. real scalar fields are described by a canonical pair consisting of a scalar 
$X$ and a conjugate scalar density of weight one $Y$ \cite{LQG1}.
\item[2.] {\it Choice of Weyl algebra}\\
The canonical quantisation requires as an input a $\ast-$algebra which is 
conveniently chosen of Weyl algebra type. In LQG this Weyl algebra is generated 
by matrix elements of holonomies of $A$ along 1-dimensional paths and exponentials 
of the flux of $E$ through 2-dimensional surfaces \cite{LQG2}. Similarly the scalar Weyl algebra
is generated by exponentials of $X$ evaluated at points and exponentials of $Y$ 
integrated over 3-dimensional regions. Note that this Weyl-algebra is manifestly background 
independent and therefore transforms covariantly under the automorphisms labelled by 
the spatial diffeomorphisms of the hypersurface.
\item[3.] {\it Choice of representations of the canonical commutation and $\ast$ relations}\\
There is a non-perturbative and background independent kinematical Hilbert space representation of 
the canonical commutation and $\ast$ relations of the Weyl algebra \cite{LQG3}. It is the unique cyclic 
representation which is invariant under the afore mentioned group of automorphisms \cite{LQG4}. 
It has a measure theoretic description and an explicit orthonormal basis (ONB)\cite{LQG5}. The 
Hilbert space is {\it not separable} and the representation is {\it irregular}, that is,
not all one parameter groups of Weyl elements are weakly continuous.   
\item[4.] {\it Solutions of Gauss and spatial diffeomorphism constraints}\\
The Gauss constraints and spatially diffeomorphism constraints respectively can be 
formulated in terms of self-adjoint respectively unitary operators on the kinematical Hilbert 
space. Their algebra is anomaly free.  
The kernel of the Gauss constraint is known explicitly, it is in fact a subspace of 
the kinematical Hilbert space. The kernel of the spatial diffeomorphism constraint 
is no longer a subspace but a space of distributions which is known explicitly \cite{LQG6}. There exist 
{\it infinitely many} 
Hilbert space structures on the space of those distributions. 
\item[5.] {\it Solutions of the Hamiltonian constraint}\\
There exist {\it infinitely many} quantisations of the Hamiltonian constraints defined on a dense invariant 
domain on the kinematical Hilbert space \cite{LQG7}. This is only possible in this representation when the density weight
of the Hamiltonian constraint equals unity. They commute with the Gauss constraint operators 
and transform as the classical counterpart under spatial diffeomorphisms up to an {\it anomalous} 
term which however annihilates spatially diffeomorphism invariant distributions. Likewise,
the commutator of two Hamiltonian constraints vanishes up to an {\it anomalous} term which however 
annihilates spatially diffeomorphism invariant distributions. Constructive solution algorithms 
exist but there is no closed solution to all Hamiltonian constraints. 
\item[6.] {\it Kinematical coherent states}\\
There exist minimal uncertainty states for any chosen, at most countably infinite selection,
of holonomy and flux operators, for instance defined by a the faces of a cell complex and 
a graph dual to it \cite{LQG8}. No coherent states exist which are minimal uncertainty states for all 
holonomy and flux operators. This is because the Hilbert space is not separable and 
the number of correlation free holonomy and flux operators is uncountably infinite.  
\item[7.] {\it Kinematical geometric operators}\\
The kinematical Hilbert space defines a representation space for self-adjoint operators 
corresponding to lengths, areas and volumes of curves, surfaces and regions in 3 spatial dimensions \cite{LQG9}.
This is despite the fact that these operators are highly non-polynomial functions of the 
flux variables.
\end{itemize}
The anomalous terms may be called {\it soft anomalies} because a solution to all constraints 
must be in particular spatially diffeomorphism invariant and thus the joint kernel of 
all constraints is a subspace of the space of diffeomorphism invariant distributions. However,
what is a {\it hard anomaly} is that up to that soft anomaly the algebra of Hamiltonian constraints 
is Abelian. This is not true in the classical theory. While in the classical theory the Poisson bracket
between Hamiltonian constraints is also a linear combination of spatial diffeomorphism 
constraints with non-vanishing structure functions as coefficients, the purely 
softly anomalous structure operators that appear
in the quantum theory are not quantisations of those classical structure functions.

Tracing back the source of the anomaly we find the {\it irregularity} of the Hilbert space 
representation. It forbids to define the Lie algebra of spatial diffeomorphisms and allows only to 
implement the corresponding Lie group. However, the Hamiltonian constraints do not generate 
a Lie algebra but rather an algebroid and a representation of the algebroid needs the 
Lie algebra generators of spatial diffeomorphisms.  

Tracing back the reason for the quantisation ambiguities we also identify the {\it irregularity}
of the Hilbert space representation. It again forbids to define the connection or scalar field 
itself and only allows to implement the corresponding holonomy or Weyl element. However, the 
classical Hamiltonian constraint is formulated in terms the connection and scalar field and not 
in terms of holonomy or Weyl elements, hence its quantisation must replace these by a function of Weyl elements.
This can be done introducing a short distance regulator and there are an infinite number of choices 
of regulated functions which in the classical theory result in the classical Hamiltonian constraint 
in the limit of vanishing regulator. Removing that regulator in the quantum theory is also possible 
in an operator topology based on spatially diffeomorphism invariant distributions. In that topology
the diffeomorphism invariant characteristics of the regulator choice survive the limit process leading 
to these quantisation ambiguities.\\
\\
In the reduced phase space route the anomaly issue disappears and the construction of the 
kernel of the constraints is obsolete as the constraints are solved already classically. This 
is a tremendous simplification. However, if one now uses the above Hilbert space representations 
as representations for observables \cite{3} such as 
the resulting physical Hamiltonian, one is again faced with an infinite number of quantisation 
ambiguities due to the irregularity of the representation. 

One can now try get rid of the ambiguities, or at least downsize their number using the idea 
of {\it renormalisation} \cite{4}. Roughly one can think of the quantisation choices as choices of couplings 
and the renormalisation group flow separates those into relevant and irrelevant. If the number
of relevant couplings is finite, then the theory is predictive. 

Another possibility is to change the representation to a more regular one which avoids those 
ambiguities from the outset. Now there are not many representations known in quantum field
theory in more than three spacetime dimensions, practically these are only the (quasi-)free
ones and in particular Fock representations. Among those one typically does not consider 
those that correspond to a diffeomorphism invariant state on an associated Weyl algebra.
Such states are more familiar from Quantum Field Theory in curved spacetime (QFT in CST) 
which take as an input a classical spacetime metric as a background and as such are 
heavily background dependent and therefore the corresponding Fock state is not diffeomorphism
invariant. 

Based on the results established in \cite{6,7} we show in this paper that there exist indeed
background independent Fock representations for both matter and geometry in $D+1$ spacetime 
dimensions under certain assumptions. In order to achieve this, we have to require the
presence of $D$ real scalar degrees of freedom, uncharged under any further Yang-Mills type 
gauge transformations. These fields could either be fundamental scalars (such as the 
standard model Higgs field, an inflaton field, a dilaton field, an axion field, other dark matter fields, etc.) 
or composite fields assembled from other fundamental fields such as the weak vector bosons of the standard model 
and gravitational fields.    
How to obtain such composite scalar fields will be described in detail in our companion paper \cite{8}.
In this paper we just assume that those $D$ real scalar fields $X^I, I=1,..,D$ and their conjugate momenta 
$Y_I$ are already available to us. \\
\\
We now go through items [1.] - [7.] of the constraint quantisation route 
described above step by step and indicate the differences:
\begin{itemize}
\item[1.] {\it Choice of polarisation}\\
Instead of canonically conjugate connections and densitised triads we consider non-densitised 
$D-Bein$ fields $e$ and conjugate momenta $P$ as fundamental geometry fields. 
These are available in any spacetime dimensions and are so(D) valued 1-forms and density one valued 
vector fields respectively. In the matter sector 
we have $D$ scalar fields $X$ with density one valued scalar conjugate momenta $Y$ as fundamental fields. 
From these 
we now construct composite fields by a canonical transformation. The result of that transformation,
which does not require a background,  
are so(D) valued scalar {\it densities of weight 1/2} $\hat{e},\hat{P}$ and uncharged, real valued
scalar {\it densities of weight 1/2} $\hat{X},\hat{Y}$ respectively. 
\item[2.] {\it Choice of Weyl algebra}\\
Instead of integrating the various fields $A,E,X,Y$ over submanifolds of complementary dimensions (adding up to 3)
we integrate all composite fields $\hat{e},\hat{P},\hat{X},\hat{Y}$ over the whole hypersurface 
smeared with test functions which geometrically are so(D) valued or scalar valued scalar densities of 
weight $1/2$ as well. Then we construct Weyl elements from those in the standard way by exponentiating these 
smeared fields. 
\item[3.] {\it Choice of representations of the canonical commutation and $\ast$ relations}\\
The key point is now the following:
The reason for carrying out the canonical transformation in step [1.] is in order to equip both 
configuration and momentum degrees of freedom with the same tensor density representation under spatial 
diffeomorphisms. Such representations are labelled by $(a,b,w)$ denoting respectively the number 
of contra-variant indices, the number of co-variant indices and the density weight. The 
fundamental fields $e,P,X,Y$ transform in the representations $(0,1,0),\; (1,0,1),\; (0,0,0),\; (0,0,1)$ 
respectively which are all different. By contrast, the composite fields transform all in the same 
representation $(0,0,1/2)$. Therefore taking complex linear combinations of the fundamental 
fields would result in geometrical nonsense as the linear combination of elements of different 
representations of the diffeomorphism group does not result in an object which transforms 
in any representation of the spatial diffeomophism group. By contrast, the complex linear 
combinations of the composite fields are simply complex valued scalar densities of weight $1/2$
and still transform in the representation $(0,0,1/2)$. All we have to do now in order to obtain 
a background independent Fock representation is to define an annihilation operator which is a complex 
linear combination of composite fields with background independent, positive coefficients. 
The Fock representation is {\it regular} and the Hilbert space is {\it separable}      
\item[4.] {\it Solutions of Gauss and spatial diffeomorphism constraints}\\
Gauss and diffeomorphism constraints, when normal ordered with respect to the 
constructed Fock structure, and not only their exponentials, are self-adjoint 
operators densely and invariantly defined on the span of Fock states. Their algebra closes without anomalies. 
An infinite number of solutions of the Gauss constraints can be constructed. These are 
all distributions and no longer elements of the Fock space. On the space of those solutions 
there is a non trivial, infinite dimensional subspace which also solves the spatial diffeomorphism 
constraint. 
\item[5.] {\it Solutions of the Hamiltonian constraint}\\
In the classical theory it is equivalent to solve the fundamental density one valued Hamiltonian constraint
or its polynomial version of higher density weight which results by multiplying the fundamental constraint 
by suitable powers of the determinant of the $D-Bein$. The normal ordered form of the polynomial 
version is a quadratic form densely defined on the span of Fock states. This is sufficient 
to formulate the quantum constraint equations. Commutators between the Hamiltonian constraint
and the Gauss and spatial diffeomorphism constraint results again in a quadratic form  
on the same domain. That part of the constraint algebra is anomaly free. Since quadratic forms cannot be 
multiplied, the commutator of Hamiltonian constraints has to be regularised. In a simpler 
context one can show that there exists 
a regularisation such that the commutator closes without anomalies. The space of solutions to 
also the Hamiltonian constraint can in principle be constructed algorithmically but is not explicitly known. 
\item[6.] {\it Kinematical coherent states}\\
There exist standard minimal uncertainty states for {\it all} creation and annihilation 
operators thanks to the separability of the Hilbert space.  
\item[7.] {\it Kinematical geometric operators}\\
Using additional non-perturbative tools, also kinematical geometric operators can be defined, however not as
self-adjoint operators but rather as symmetric quadratic forms and not on the span of the 
Fock states but on another dense domain defined by a kinematical coherent state 
with respect to which the quantum metric has a non-degenerate metric as expectation value.  

\end{itemize}
Concerning the reduced phase space quantisation route, provided the reduction process 
which involves the choice of a reference frame (defined by a choice of reference fields and 
gauge fixing conditions on them) does not use any background structures, the above 
type of Fock representations can also be used to define the physical Hamiltonian as a 
quadratic form on the chosen Fock space. Since the physical Hamiltonian naturally carries 
density weight unity which makes it non-polynomial, one needs to use the tools from step 
[7.] to define it. The same applies to most other observables, in particular the geometric 
operators which after reduction are in fact promoted to gauge invariant observables. \\
\\
In what follows we outline the details of these observations in brief terms, leaving the details 
to \cite{8}. The manuscript is organised according to [1.] - [7.] as follows:\\
\\ 

In section \ref{s2} we introduce the polarisation and the canonical transformation 
to half density valued scalars of all field species. This step is mirrored in LQG by 
the chain of canonical transformations $e,P\to E,K\to A=\Gamma+\beta K, E/\beta$ where 
$\Gamma$ is the spin connection and $\beta$ the Immirzi parameter. Immirzi like 
parameters are also possible in the present formulation.

In section \ref{s3} we introduce the Weyl algebra. This step is much simpler than in LQG 
because we smear the fields in all directions and not with distributional form factors.

In section \ref{s4} we pick concrete background independent annihilation operators to 
define a Fock structure. 

In section \ref{s5} we write the Gauss and spatial diffeomorphism constraint in terms 
of annihilation and creation operators and normal order. The crucial observation is that 
the resulting expressions are bi-linear in those with no creation operator squared terms.
This means that they are self-adjoint operators on a dense invariant domain of the Fock space 
and not only symmetric quadratic forms. Therefore their commutators can be computed without 
effort and the algebra closes in the expected way. We show that there exists an infinite 
number of common generalised zero eigenvectors for both constraints.

In section \ref{s6} we remark that the Hamiltonian constraint in density weight 
six version which is obtained by multiplying the density one version with the fifth 
power of $\det(e)$ is a symmetric quadratic form densely defined on the same 
domain as the other constraints when normal ordered. Commutators 
between Gauss respectively the spatial diffeomorphism constraint can therefore be immediately meaningfully 
computed and close in the expected way. The commutator between Hamiltonian constraints is 
more delicate as quadratic forms cannot be multiplied, but we argue that the computation 
can be regulated and results a normal ordered quantisation of the classical counterpart.
We will not carry out that calculation in the present paper as the density six valued 
Hamiltonian constraint is a tenth order polynomial but we refer the reader to \cite{7} where 
that computation was carried out successfully for pure Euclidian gravity in selfdual variables 
in its fourth order polynomial density two valued version.  
We also point out that quadratic forms are sufficient to determine constraint solutions and 
suggest an algorithm for obtaining them.

In section \ref{s7} we define the standard coherent states available for any Fock representation 
which in our case are minimal uncertainty states for the half density valued scalars. 

In section \ref{s8} we construct geometrical operators. These are non-polynomial in the 
the annihilation and creation operators and thus can only be defined as symmetric quadratic 
forms on the Fock space on a dense domain that is defined by coherent states which 
fluctuate around {\it non-degenerate} classical metrics. Their generalised eigenvalues 
are {\it not discrete}.

In section \ref{s9} we truncate the theory to the cosmological sector. We show that 
by exploiting the ordering ambiguity in the quantisation of a non-linear Hamiltonian it 
is possible to motivate a Hamiltonian that predicts a cosmological bounce despite the 
fact that we use a regular representation on a separable Hilbert space.

In section \ref{s10} we summarise and conclude.

\section{Polarisation and half density valued scalars}
\label{s2}

In the Palatini polarisation, after solving second class constraints associated with the 
boost part of the gravitational so(1,D) Gauss constraint, the gravitational field sector of the phase space 
is coordinatised by a canonical pair $(e_a^j,p^a_j)$ where $a,b,c,..=1,..,D$ are spatial
tensor indices with respect to a spatial manifold $\sigma$ 
while $j,k,l,..=1,..,D$ are so(D) indices. The scalar field field sector of the phase space 
is coordinatised by a canonical pair $(X^I,Y_I)$ with $I,J,K,..=1,..,D$. All fields are 
real valued. 

As the spatial metric $q_{ab}=\delta_{jk} e^j e ^k_b$ is supposed to be non-degenerate
one assumes $\det(e)\not=0$. This anholonomic constraint on the classical phase space is 
ignored upon quantisation. In the connection polarisation it translates into $\det(E)\not=0$ and 
that classical constraint is drastically violated in LQG as spin network quantum states 
have zero volume everywhere except for a finite number of points. The argument is that 
the anholonomic constraint is a condition that is part of the definition of a semiclassical state 
and we adopt that viewpoint.

From the scalar fields we can construct a flat co-D-Bein
\be \label{2.1}
h^I_a:= X^I_{,a}
\ee
with associated flat metric $h_{ab}:= \delta_{IJ} h^I_a h^J_b$. Just as for $q$ we 
require that $h$ is non-degenerate in the classical theory and lift that condition in 
the quantum theory. Thus we have now, in the classical theory, a bi-metric theory 
where both $q,h$ have Euclidian signature. We denote the corresponding 
D-Beine by $e^a_j, h^a_I$ where $e^a_j e_a^k=\delta_j^k$ and  $h^a_I h_a^J=\delta_I^J$.

With these preparations we can now pass to the half density valued scalar formulation 
via a chain of three simple canonical transformations:
\begin{itemize}
\item[i.] {\it Removing the spatial indices}\\
We define 
\be \label{2.2}
\tilde{e}_I^j:=h^a_I\; e_a^j,\; \tilde{p}^I_j:=h_a^I \; p^a_j,\; 
\tilde{X}^I:=X^I,\; \tilde{Y}_I:=Y_I-(p^a_j e_b^j h^b_I)_{,a}
\ee
% P de =h^{-1} \tilde{P} d(h \tilde{e})=\tilde{P} d\tilde{e} + h^{-1} \tilde{P} \tilde{e} dh
% =\tilde{P} d\tilde{e} +P e h^{-1} dh
Since $p$ is a vector density we see that $\tilde{e},\tilde{X}$ are scalars of density weight 
zero and $\tilde{p},\tilde{Y}$ are scalars of density weight one. It is easy to check 
that (\ref{2.2}) is a canonical transformation with inverse 
\be \label{2.3} 
e_a^j=\tilde{h}_a^I\; \tilde{e}_I^j,\; p^a_j=\tilde{h}^a_I \; \tilde{p}^I_j,\; 
X^I=\tilde{X}^I,\; Y_I=\tilde{Y}_I+(\tilde{h}^a_J\; \tilde{p}^J_j \tilde{e}_I^j)_{,a} 
\ee
where we have abbreviated $\tilde{h}_a^I:=\tilde{X}^I_{,a}$ with inverse $\tilde{h}^a_I$.
\item[ii.] {\it Distributing the density weight equally in the geometry sector}\\
We define 
\be \label{2.4}
\check{e}_I^j:=\tilde{e}_I^j |\det(\tilde{h})|^{1/2},\; \check{p}^I_j:=\tilde{p}^I_j\; |\det(\tilde{h})|^{-1/2},\;
\check{X}^I:=\tilde{X}^I,\; \check{Y}_I:=\tilde{Y}_I+\frac{1}{2}(\tilde{p}^J_j \tilde{e}_J^j \tilde{h}^a_I)_{,a}
\ee
% H=|det(h)|^{1/2}
% P de = H P' d(H^{-1} e')= P' de' - H^{-1} P' e' dH= P' de' - P e Tr(h^{-1/2} partial h)/2 
%
Since $|\det(\tilde{h})|^{1/2}$ has density weight $1/2$, it follows that both $\check{e},\check{p}$ 
are scalars of density weight $1/2$ while $\check{X}, \check{Y}$ are scalars of density weight 
zero and one respectively.
It is easy to check 
that (\ref{2.4}) is a canonical transformation with inverse 
\be \label{2.5}
\tilde{e}_I^j=\check{e}_I^j |\det(\check{h})|^{-1/2},\; \tilde{p}^I_j=\check{p}^I_j\; |\det(\check{h})|^{1/2},\;
\tilde{X}^I=\check{X}^I,\; \tilde{Y}_I=\check{Y}_I-\frac{1}{2}(\check{p}^J_j \check{e}_J^j \check{h}^a_I)_{,a}
\ee
where we have abbreviated $\check{h}_a^I:=\check{X}^I_{,a}$ with inverse $\check{h}^a_I$.
\item[iii.] {\it Distributing the density weight equally in the matter sector}\\
We define 
\be \label{2.6}
\hat{e}_I^j:=\check{e}_I^j, \; \hat{p}^I_j:=\check{p}^I_j- \frac{1}{D}\tilde{Y}^J \tilde{X}_J \tilde{e}^I_j,\; 
\hat{X}^I:=\check{X}^I |\det(\check{e})|^{1/D},\; 
\hat{Y}^I:=\check{Y}_I |\det(\check{e})|^{-1/D}
\ee
% E=det(e)^{1/D}
% Y dX= E Y' d(X'/E)=Y' dX'-Y' X' dE/E=Y' dX' -Y X 1/D Tr( e^{-1} de)
where $\tilde{e}^I_j$ is the inverse of $\tilde{e}_I^j$. 
Since $|\det(\tilde{e})|$ carries density weight $D/2$ it follows that now 
all fields carry density weight $1/2$. It is easy to check 
that (\ref{2.6}) is a canonical transformation with inverse 
\be \label{2.7}
\check{e}_I^j=\hat{e}_I^j, \; \check{p}^I_j=\hat{p}^I_j+\frac{1}{D}\hat{Y}^J \hat{X}_J \hat{e}^I_j,\; 
\check{X}^I=\hat{X}^I |\det(\hat{e})|^{-1/D},\; 
\check{Y}^I=\hat{Y}_I |\det(\hat{e})|^{1/D} 
\ee
where $\hat{e}^I_j$ is the inverse of $\hat{e}_I^j$.
\end{itemize}
Note that the transformation i.-iii. is local but not ultra-local, it involves partial 
derivatives. The reason why it is not necessary to solve partial differential equations (PDEs)
to construct the inverse is that in each step the field whose derivatives are involved 
itself transforms trivially. This would be different if one tried to define scalar densities
of weight 1/2 just using geometry fields: This would necessarily involve scalar curvature 
polynomials and such a transformation would require to solve PDEs which is practically impossible
without using perturbative/iteration methods and would lead to extremely non-local expressions.

Finally note that we can still allow for additional canonical transformations such as 
the "Immirzi" like rescalings $\hat{e},\hat{p}\to \gamma \hat{e},\gamma^{-1} \hat{p},\; \gamma\not=0$ and similar for 
$\hat{X},\hat{Y}$. We will not enter into those generalisations to keep the presentation simple.

\section{Weyl algebra}
\label{s3}

The definition of the Weyl algebra is extremely simple: Let 
$f^I_j,\; g_I^j,\;F_I,\; G^I$ be test functions on $\sigma$, say smooth and of rapid decrease or 
of compact support. Note that these notions do not involve a background metric.
Then the Weyl elements are 
defined by 
\be \label{3.1}
W(f,g;F,G)=\exp(i\int_\sigma\; d^Dx\; [f^I_j \; \hat{e}_I^J+g_I^j \; \hat{p}^I_j+ F_I\; \hat{X}^I+ G^I\; \hat{Y}_I])   
\ee
Geometrically $f,g$ are so(D) valued scalar densities of weight 1/2 while $F,G$ are just real valued 
scalar densities of weight 1/2. The Weyl relations are the usual ones 
\ba \label{3.2}
W(f,g;F,G)^\ast &=& W(-f,-g;-F,-G)
\\
W(f,g;F,G)\; W(f',g';F',G') &=& e^{i[<f,g'>-<f',g'>+<F,G'>-<F',G>]}\; W(f+f',g+g',F+F',G+G')
\nonumber
\ea
where we have defined the scalar product of (copies of) the one particle Hilbert space $\mathfrak{h}=L_2(\sigma,d^Dx)$
\be \label{3.3}
<f,g>= \int_\sigma\; d^Dx\; \overline{f^I_j}\; g_I^j,\; <F,G>= \int_\sigma\; d^Dx\; \overline{F_I}\; G^I
\ee
(the test functions in (\ref{3.1}) are of course real valued). Note that this inner product is 
background independent because the test functions are half density valued. It is also positive 
definite as we raise and lower the indices $I,j$ with the respective Kronecker symbol.
These Kronecker symbols are no background metrics on $\sigma$ but rather arise as the Cartan-Killing 
metric on so(D) and on the internal scalar field species space.  

The Weyl algebra is now the abstract $^\ast-$algebra $\mathfrak{A}$ generated by the Weyl elements (\ref{3.1}) 
subject to the Weyl relations (\ref{3.2}). Note how simple this definition is as compared to the 
holonomy flux algebra which requires a substantial amount of differential geometry to define the notion of 
semi-analytic structures.

\section{Fock state and Fock representation}
\label{s4}

The Fock representation is the Hilbert space representation of $\mathfrak{A}$ which arises 
as the Gel'fand-Naimark-Segal (GNS) data $(\pi,{\cal H},\Omega)$ from a Fock state, i.e. a positive, 
linear, normalised functional $\omega$ on $\mathfrak{A}$. It is completely specified by 
its action on Weyl elements. We pick the following Fock state: 
\be \label{4.1}
\omega(W(f,g;F,G)):=e^{-\frac{1}{4}[<f,f>+<g,g>+<F,F>+<G,G>]}
% exp(i(bq+ap))=exp(i[b(z+z*))+ai(z-z*)]/sqrt(2))=exp(i[(b+ia)z+(b-ia) z*)]/sqrt(2))
% exp(i[ cz+c* z*)]/sqrt(2))=exp(i c* z*/sqrt(2)) exp(i cz/sqrt(2)) \exp(-1/2[i c^* z*,ic z]/sqrt(2)^2)
% =exp(-|c|^2/4) 
\ee
One obtains (\ref{4.1}) by defining the annihilation operators 
\be \label{4.2}
a_I^j:=2^{-1/2}[\hat{e}_I^j-i\; \delta_{IJ}\delta^{jk}\; \hat{p}^J_k],\;
A_I:=2^{-1/2}[\delta_{IJ }\hat{X}^J-i\; \hat{Y}^I]
\ee
and the cyclic GNS vector $\Omega$ is the Fock vacuum $a_I^j \Omega=A_I\Omega=0$. The non-trivial 
canonical commutation relations (CCR) are
\be \label{4.2a}
[a_J^j(x),(a_K^j(y))^\ast]=\delta_{JK}\delta^{jk}\delta(x,y),\; 
[A_J(x),(A_K(y))^\ast]=\delta_{JK}\delta(x,y)
\ee
The state (\ref{4.1}) is the analog of the state in LQG that defines a cyclic representation of the 
holonomy flux algebra and which defines an orthonormal basis given by spin network functions. 
It can be defined by 
\be \label{4.3}
\omega_{{\sf LQG}}(T_s\; W)=\delta_{s,0}
\ee
where $T_s$ is a spin network multiplication operator and $W$ the product of exponentials 
of fluxes. The $T_s\; W$ are the analogs of the Weyl elements. The definition 
(\ref{4.3}) displays the drastic discontinuity of the LQG representation with respect 
to changes of the label $s$ which involves graphs, spin representation labels and intertwiner 
labels. By contrast, the state (\ref{4.1}) is entirely continuous with respect to 
all smearing functions. Therefore the operator valued distributions $\hat{e},\hat{p},\hat{X},\hat{Y}$ 
exist in this representation in contrast to the connection $A$ in the LQG representation.
Smeared versions of the $\hat{e},\hat{p},\hat{X},\hat{Y}$ define self-adjoint operators by Stone's 
theorem because regularity of $\omega$ implies weak continuity of 1-parameter unitary groups of 
Weyl elements. 

Note also that the state $\omega$ is both Gauss invariant and spatially diffeomorphism invariant thanks to 
the manifest Gauss invariance and spatial diffeormorphism invariance of the 1-particle inner product
(\ref{3.3}). More specifically, we have automorphisms $\alpha_r,\beta_\varphi$ with 
local rotations $r:\sigma\to$SO(D)) and diffeomorphisms $\varphi\in$Diff$(\sigma)$ acting on $\mathfrak{A}$ as 
\be \label{4.4}
\alpha_r(W(f,g;F,G))=W({\sf Ad}_{r^{-1}}(f), {\sf Ad}_{r^{-1}}(g);F,G),\;\;
\beta_\varphi(W(f,g;F,G))=W((\varphi^{-1})^\ast(f,g;F,G))
\ee
Here Ad denotes the adjoint action of the group on the Lie algebra and $\varphi^\ast$ is the 
pull-back map.
These are automorphism groups $\alpha_r\circ\alpha_{r'}=\alpha_{r'\cdot r},\;
\beta_\varphi\circ\beta_{\varphi'}=\beta_{\varphi'\circ \varphi}$ and the state is 
invariant 
\be \label{4.5}
\omega\circ \alpha_r=\omega\circ \beta_\varphi=\omega
\ee
Thus the operators densely defined by 
\be \label{4.6} 
U(r)\pi(W)\Omega:=\pi(\alpha_r(W))\Omega,\;V(\varphi)\pi(W)\Omega:=\pi(\beta_\varphi(W))\Omega,\;
\ee
for $W\in \mathfrak{A}$ extend to unitary operators on the Fock space. Again thanks to the 
regularity of the state, one parameter unitary subgroups of SO(D) and Diff$(\sigma)$ 
act weakly continuously which is why we can define self-adjoint generators known as 
Gauss and spatial diffeomorphism constraints. These will be subject of the next section.

Next note that the Fock space is separable. Indeed a countable orthonormal basis (ONB) can be 
defined as follows: Pick a real test function valued ONB $b_n$ of the 1-particle Hilbert space
$\mathfrak{h}$ called modes and define for $N_{n,J}^j,\; M_{n,I}\in \mathbb{N}_0$ 
\be \label{4.7}
B_{M,N}:=\prod_{m,n,I,J,j}\; [\frac{(A_{m,I}^\ast)^{M_{m,I}}}{\sqrt{M_{m,I}!}} \; \frac{((a_{n,J}^j)^\ast)^{N_{n,J}^j}}{\sqrt{N_{n,J}^j!}}] \; \Omega
\ee
where 
\be \label{4.8}
a_{n,J}^j:=<b_n,a_J^j>, \; A_{n,I}:=<b_n,A_I>
\ee
Then the $B_{M,N}$ with $M_{m,I},N_{n,J}^j\not=0$ for finitely many $m,I,n,J,j$ provide an ONB of the Fock space.

It is also well known that the the Fock representation is irreducible for the Weyl algebra. 
This means that every vector is cyclic. Formally this follows from the fact that the span 
of the Fock states $B_{M,N}$ is dense, that one can reach any $B_{M',N'}$ from 
any $B_{M,N}$ by a monomial in creation and annihilation operators and that the latter 
operators can be obtained as strong limits of Weyl elements. 
Moreover, the GNS Hilbert space can be displayed as a rigorous $L_2$ space with respect 
to a Gaussian measure supported on tempered distributions. This measure theoretic 
presentation is the analog of the Ashtekar-Lewandowski measure in LQG.  

Finally a remark on uniqueness:
We still obtain a Gauss and spatially diffeomorphism invariant state as long as the exponent 
in (\ref{4.1}) remains invariant under the adjoint and pull-back map. This essentially 
confines the metric on the internal so(D) space to be the Kronecker metric while 
we can replace the Kronecker metric on the internal field species indices $I,J$ by any other 
{\it constant} positive definite metric. Thus the invariant Fock state is essentially unique 
up to a finite parameter set of possibilities. We will not enter into those extensions to 
keep the presentation simple.

\section{Gauss and spatial diffeomorphism constraints}
\label{s5}

In principle we can define the Gauss and spatial diffeomorphism constraint as the 
self-adjoint generators $G[l],\;D(u)$ of the 1-parameter unitary groups 
\be \label{5.1}
e^{i\;s\;G(l)}:=U(e^{s\; l}),\; 
e^{i\; t D(u)}:=V(\varphi^u_t)
\ee
defined in the previous section where $l:\sigma\to$so(D) is a Lie algebra valued function and 
$\varphi^u_t$ the 1-parameter group of diffeomorphisms defined by the integral curves of 
the vector field $u$ on $\sigma$. However, it is instructive to derive explicit expressions 
in terms of the annihilation and creation operators.

The constraints in terms of the original canonical variables $e,p,X,Y$ read
\be \label{5.2}
G(l)=\int\; d^Dx\; l_{jk}\; e_a^j \, p^a_i \delta^{ki},\;\;
D(u)=\int\; d^Dx\; [p^a_j\; ({\cal L}_u \;e^j)_a+Y_I\;({\cal L}_u \; X^I)_a]
\ee
where $l_{(jk)}=0$ is antisymmetric and $\cal L$ denotes the Lie derivative.

After the canonical transformation (\ref{2.2}), (\ref{2.4}), (\ref{2.6}) the constraints
read 
\be \label{5.2a}
G(l)=\int\; d^Dx\; l_{jk}\; \hat{e}_I^j \, \hat{p}^I_i \delta^{ki},\;\;
D(u)=\int\; d^Dx\; [\hat{p}^I_j\; ({\cal L}_u \;\hat{e}_I^j)_a+\hat{Y}_I\;({\cal L}_u \; \hat{X}^I)_a]
\ee
as one can explicitly check but which also follows immediately from the geometrical 
nature of (\ref{2.2}), (\ref{2.4}), (\ref{2.6}). Note that $D(u)$ now contains the 
Lie derivative of half density valued scalars $S$ only which reads explicitly 
\be \label{5.3}
{\cal L}_u S =u^a \; S_{,a}+\frac{1}{2} u^a_{,a} S
\ee
We integrate the $u^a_{,a}$ term by parts and obtain 
\be \label{5.4}
D(u)=\frac{1}{2}
\int\; d^Dx\; u^a\; [\hat{p}^I_j\;\hat{e}_I^j)_{,a} -(\hat{p}^I_j)_{,a}\;\hat{e}_I^j
+\hat{Y}_I\;\hat{X}^I_{,a}-\hat{Y}_{I,a}\;\hat{X}^I]
\ee 
where we dropped the boundary term using that $u$ decays sufficiently fast which is due to
$u$ defining a gauge transformation (in contrast to a symmetry transformation).  We now express 
(5.4) in terms of annihilation and creation operators and normal order
\ba \label{5.5}
G(l) &=& -i\int\; d^Dx\; l_{jk}\; \delta^{JK}\;(a_J^j)^\ast \; a_K^k
%i/2Tr(l((a+a*)(a-a*)))=-iTr(a* a)
\nonumber\\
D(u) &=& -\frac{i}{2}
\int\; d^Dx\; u^b\; \delta^{JK}\{\delta_{jk}\;[(a_J^j)^\ast \;(a_K^k)_{,b}- (a_J^j)^\ast_{,b} \;a_K^k]
+[A_J^\ast (A_K)_{,b}-(A_J^\ast)_{,b}\; A_K]\}   
%i/4 Tr((a-a*)(a+a*)'-(a-a*)'(a+a*)=-i((a*) a'-(a*)' a)/2 
\ea
Formula (\ref{5.5}) reveals the crucial point: Due to all fields having density weight 1/2, both constraints 
do not contain terms of the form $(a^\ast)^2, \; (A^\ast)^2$. Such terms typically make the 
quantum object no longer an operator but only a quadratic form unless delicate Hilbert-Schmidt conditions 
are satisfied for the coefficients of such terms. By contrast, (\ref{5.5}) is densely and invariantly 
defined on the span of Fock states. Since they generate the subgroups (\ref{5.1}) they are densely defined on 
the domain on which (\ref{5.1}) is strongly continuous by Stone's theorem. Commutators can easily 
be computed and close in the expected anomaly-free way 
\be \label{5.6}
[G(l),G(l')] = -i G([l,l']), \; \;[D(u),G(l)] = -i G(u[l]),\;\;[D(u),D(u')] = -i D([u,u'])
\ee
as one can verify by straightforward computation in a few lines just using the CCR (\ref{4.2a}). Here 
$u[l], [u,u']$ denote the action of vector fields $u$ on scalars $l$ and on vector fields $u'$ (by commutator) respectively.

We define a generalised zero eigenvector to both constraints to be a linear functional 
$L:\; {\cal H}\to \mathbb{C}$ such that 
\be \label{5.7}
L[G(l)\psi]=L[D(u)\psi]=0
\ee
for all $l,u$ and $\psi\in {\cal D}$ where $\cal D$ is the span of Fock states. An infinite number 
of solutions to (\ref{5.6}) can be constructed as follows: Define the pair creators 
\be \label{5.8}
c_{JK}:=\int\; d^Dx\;\delta_{jk}\; (a_J^j)^\ast\; (a_K^k)^\ast,\; 
C_{JK}:=\int\; d^Dx\; (a_J)^\ast\; (a_K)^\ast
\ee
These are no longer operators on Fock space but only quadratic forms with domain $\cal D$. 
However, they are easily verified to commute with (\ref{5.5}) by explicit computation 
or by inspection because $G(l)$ rotates 
the $a$ while leaving the $A$ invariant and because $[D(u),A_J^\ast A_K^\ast]=-i(u^a A_J^\ast A_K^\ast)_{,a}$ is 
a total derivative and similar for $a$. It follows that the linear functionals  
\be \label{5.9}
L:= <{\sf Pol}(C,c)\Omega,\;>
\ee
solve (\ref{5.7}) where Pol is any polynomial in the quadratic forms (\ref{5.8}) because 
$G(l)\Omega=D(u)\Omega=0$. 

Note that 
none of the ``vectors'' ${\sf Pol}(C,c)\Omega$ except $\Omega$ is normalisable, these define
therefore generalised zero eigenvectors. As compared to LQG we do not have the complete 
set of solutions to the Gauss constraint but on the other hand we have a more controlled 
set of solutions to the spatial diffeomorphism constraint as in LQG those solutions 
involve an uncountably infinite sum over spin network functions over the same diffeomorphism 
equivalence class of graphs.

\section{Hamiltonian constraint}
\label{s6}

The Hamiltonian constraint in the original variables $e,p,X,Y$ and density weight one version is given by 
\be \label{6.1}
C(f)=\int\; d^Dx\;f \{[\frac{1}{Q}(q_{ac} q_{bd}-\frac{1}{D-1} q_{ab} q_{cd}) p^{ab} p^{cd}-Q\;R[q]+\Lambda \;Q]
+\frac{1}{2}[\frac{\delta^{JK} Y_J Y_K}{Q}+Q\;( \delta_{JK} q^{ab} X^J_{,a} X^K_{,b}+W(X))]\}
\ee
where $R[q]$ is the Ricci scalar of $q$, $W$ is a polynomial scalar potential and the notation is as follows:
\be \label{6.2}
q_{ab}=\delta_{jk} e^j_a e^k_b, \; q^{ac} q_{cb}=\delta^a_b,\; Q=\sqrt{\det(q)},\; 2\; p^{ab}= p^{(a}_j e^{b)}_k \delta^{jk}
\ee
We could generalise (\ref{6.1}) by replacing the Kronecker matrices with respect to the scalar field species by a positive 
definite constant matrix and its inverse but by principal axis transformation we would get back to (\ref{6.1}).   

We must write (\ref{6.1}) in terms of $\hat{e},\hat{p},\hat{X},\hat{Y}$. This will lead to a rather complicated 
expression which will be further detailed in \cite{8}. However, what is clear from the form of the transformations 
(\ref{2.2}), (\ref{2.4}) and (\ref{2.6}) is that, without going into these details,
upon multiplying $C$ by suitable powers of $\det(e), \det(h)$ we can write the so rescaled 
$\hat{C}$, which is a scalar density of some density weight $w$, 
as a polynomial in $\hat{e},\hat{p},\hat{X},\hat{Y}$. (The details not displayed here involve 
using a Gauss invariant, half density valued  geometrical function different from $|\det(\check{e})|^{1/D}$ 
in step iii. of section \ref{s2} which is necessary to reach a polynomial form).

We can then use (\ref{4.2}) to write it as 
a polynomial in $a,a^\ast, A, A^\ast$ and normal order. The resulting expression, denoted as $\hat{C}(\hat{f})$ 
where we distinguish between the scalar smearing function $f,\hat{f}$ of density weight zero and  
$ 1-w$ respectively, is then 
a well defined symmetric quadratic form with dense domain $\cal D$ but no longer an operator because it contains many 
terms with more that one creation operator factor. This however is sufficient in order 
to define solutions to the quantum constraint equations in analogy to (\ref{5.6}) as 
\be \label{6.3}
L(\hat{C}(\hat{f})\psi)=0
\ee
for all $\hat{f}$ and $\psi\in {\cal D}$. To solve (\ref{6.3}) one may proceed as follows: The linear 
functional $L$ is completely specified by its action on the basis elements $B_{M,N}$ (\ref{4.7}). 
Thus 
\be \label{6.4}
L=\sum_{M,N}\; L_{M,N}\; <B_{M,N},.>,\; L_{M,N}=L(B_{M,N})
\ee
Then, also decomposing $\hat{f}=\sum_n \hat{f}_n b_n$, (\ref{6.4}) is equivalent to the infinite system of linear equations 
\be \label{6.5}
\sum_{M,N}\; L_{M,N}\; <B_{M,N}, \hat{C}(b_n)\;B_{M',N'}>=0
\ee
for all $n,M',N'$. Since the matrix elements $<B_{M,N}, \hat{C}(b_n)\;B_{M',N'}>$ of a quadratic form all 
exist, (\ref{6.5}) can be rigorously defined. To actually solve the system depends on the details for 
which we refer to \cite{8}. The system (\ref{6.5}) is the analog of the system in LQG where the 
$B_{M,N},B_{M',N'}$ are replaced by the uncountable basis of spin network functions $T_s,T_{s'}$ 
and the label $n$ is replaced by 
the vertices of the graph underlying $T_{s'}$.

Finally, we consider the quantum constraint algebra. The integrand of $\hat{C}(f)$ is a manifestly rotation 
invariant scalar density of some density weight $w$. Therefore it is clear without any computation 
that \be \label{6.6}
[G(l),\hat{C}(\hat{f})]=0,\; [D(u),\hat{C}(f)]=-i\hat{C}(u[\hat{f}])
\ee
where $u[\hat{f}]$ is the Lie derivative with respect to $u$ of the scalar density $\hat{f}$ of weight $1-w$. Hence 
there are no anomalies from the Gauss and spatial diffeomorphism constraint. Actually (\ref{6.6})
has to be understood in the weak operator topology: We sandwich both sides between elements of 
$\cal D$, then as $G(l),D(u)$ preserve $\cal D$, the matrix elements are well defined.  

The commutator between two Hamiltonian constraints however is delicate: As quadratic forms 
cannot be multiplied, the commutator is ill-defined as it stands. Thus it must be regularised. 
This can be done using a truncation: We decompose $\hat{C}(\hat{f})$ with respect to the 
(\ref{4.8}) and keep only the terms labelled by modes $n\le N$ for some $N$. The resulting 
truncated objects $\hat{C}_N(\hat{f})$ are then operators densely and invariantly defined on $\cal D$. One can 
compute the commutator, say $[\hat{C}_N(\hat{f}),\hat{C}_{N'}(\hat{f}')]$ which is well defined,
normal order the result and take suitable limits $N,N'\to \infty$. If the terms that in the naive 
commutator of quadratic forms computation diverge can be made to cancel in that process then 
the result of the computation must be 
\be \label{6.7}
[\hat{C}(\hat{f}),\hat{C}(\hat{f}')]=-i\; :\{\hat{C}_N(\hat{f}),\hat{C}_{N'}(\hat{f}')\}:
\ee
where the right hand side denotes the normal ordered version of the classical computation.
To see how this works in principle in a much simpler setting, see \cite{7}. For the 
status of this computation for the present case, see \cite{8}.

\section{Kinematical coherent states}
\label{s7}

These are simply the normalised standard states 
\be \label{7.1}
\Omega_{z,Z}:=e^{-(||z||^2+||Z||^2)/2}\; e^{<z,a>^\ast+<Z,A>^\ast}\;\Omega
\ee
where the norms and scalar products refer to the one particle Hilbert space $\mathfrak{h}\ni z,Z$.
These are minimal uncertainty states for $\hat{e},\hat{p},\hat{X},\hat{Y}$ and expectation 
values given by real and imaginary parts of of $z,Z$ respectively. In contrast to the 
kinematical coherent states for LQG which display semiclassical properties only for 
a countable collection of holonomies and fluxes but have uncontrollable expectation value and 
large fluctuation for 
an uncountably infinite number of these, the coherent states (\ref{7.1}) have semiclassical properties 
for {\it all} $a_{n,J}^j, A_{n,J}$ and their adjoints. The reason for this much improved 
semiclassical sector is the {\it separability} of the Fock space in contrast to the {\it non-separability}
of the LQG Hilbert space.

\section{Kinematical geometric operators}
\label{s8}

As geometrical kinematical functions corresponding to volumes of submanifolds of $\sigma$
of dimension $1,2,..,D$ are non-polynomial functions of $\hat{e},\hat{X}$ we will not be able 
to implement these as operators but only as quadratic forms. For general dimension of the 
submanifold see \cite{8}. In this paper we consider the simplest case, the $\hat{e}$ volume of 
a D-dimensional region $R$ whose classical expression is 
\be \label{8.1}
{\sf Vol}(R):=\int_R \; d^Dx\; |\det(\hat{e})|^{2/D}    
\ee
A direct expression of $\hat{e}$ in terms of annihilation and creation operators does 
not lead to a meaningful object on Fock space. Instead we invoke methods from 
deformation quantisation and note the following identity
\be \label{8.2} 
|\det(\hat{e}(x))|^{-(D+r)}=\frac{f_r(\hat{e}(x))}{f_r(1_D)},\; 
f_r(M):=\int_{\mathbb{R}^{D^2}}\;d^{D^2}u\; |\det(u)|^r \; f(M\cdot u)   
%u'=u\cdot : u'_I=u_j e^j_I: D times gives J^{-D} from measure and J^{-r} from determinant 
%where J is the Jacobean of e, altogether $J^{-(D+r)} 
\ee
for any $r\ge 0$. Here $f:\; \mathbb{R}^{D^2}\to \mathbb{R}$ is a function 
mapping real valued matrices in $D$ dimensions to real numbers such that the integral $f_r(1_D)$
converges for any $r\ge 0$. An example would be the Gaussian function $f(u)=\exp(-{\sf Tr}(u^T \;u))$.
Then we have 
\be \label{8.3}
|\det(\hat{e}(x))|^{2/D}=[\det(\hat{e}(x))]^{2k} \; \frac{f_{2k-D-2/D}(\hat{e}(x))}{f_{2k-D-2/D}(1_D)}
%2k-(D+r)=2/D, 2k=D+r+2/D even integer r=2k-D-2/D \ge 0 
\ee
for any positive integer $k$ such that $r=2k-D-2/D\ge 0$. For $D=3$ the minimal $k$ would be $k=2$. 

Let $\hat{f}$ be the Fourier transform of $f$. Then we may define (\ref{8.1}) as the quadratic 
form on the Fock space by 
\be \label{8.4}
{\sf Vol}(R):=\frac{1}{f_{2k-D-2/D}(1_D)}\int\; d^{D^2}u \; |\det(u)|^{2k-D-2/D}
\int_R \; d^Dx\; \int\; \frac{d^{D^2}p}{(2\pi)^{D^2}}\; 
:\;\det(\hat{e}(x))^{2k} e^{i{\sf Tr}(p\cdot \hat{e}(x)\cdot u)}\;:     
%p^\alpha_I e_I^j u^j_\alpha
\ee
To carry out the normal ordering involved we write out $\det(\hat{e})$ as a polynomial in annihilation 
and creation operators and we have 
\be \label{8.5}
:\;e^{i {\sf Tr}(p\cdot \hat{e}\cdot u)}\; : = e^{i {\sf Tr}(p\cdot a^\ast\cdot u)}\; 
 e^{i {\sf Tr}(p\cdot a \cdot u)}
\ee
Then it is easy to show, using the eigenvector property of coherent states for the annihilation 
operators, that 
\be \label{8.6}
<\Omega_{z,Z},\; {\sf Vol}(R)\; \Omega_{z,Z}> = {\sf Vol}_z(R)
\ee
where ${\sf Vol}_z(R)$ is the classical volume (\ref{8.1}) for $\hat{e}=(z+\bar{z})/\sqrt{2}$. 

A possible dense domain of definition of (\ref{8.4}) is therefore the span of excitations
of a coherent state $\Omega_{z,Z}$ such that $(z+\bar{z})(x)/\sqrt{2}$ is nowhere degenerate.
Here the excitations of $\Omega_{z,Z}$ are the polynomials in creation operators applied to 
it. That space of states so obtained is dense is due to the fact that the Fock representation 
is irreducible, i.e. every vector is cyclic.

It is interesting to see that the anholonomic constraints $\det(e),\det(h)\not=0$ of the
classical theory which imply $\det(\hat{e})\not=0$ find their way into domain conditions 
in the quantum theory. The tools that we have used above in order to define non-polynomial 
objects on Fock space as quadratic forms are the direct analog of the inverse volume techniques 
used in LQG to define inverse volume and curvature operators.

\section{Cosmological truncation}
\label{s9}

In the cosmological truncation of LQG called Loop Quantum Cosmology (LQC) one takes the 
full LQG theory as the motivation to introduce a rather unusual Hilbert space representation 
of the Weyl algebra. It mimics several features of the full theory, in particular the 
irregularity of some of the Weyl elements and the non-separability of the Hilbert space. 
The irregularity of some of the Weyl elements leads to an unusual quantisation (called polymerisation) 
of the cosmological 
Hamiltonian which effectively turns a classically unbounded function into a bounded 
operator. Likewise, the regular Weyl elements are turned into operators with discrete rather 
than continuous spectrum. The pay-off of such a quantisation is that e.g. inverse volume analogs are 
bounded from below leading to a cosmological bounce. The drawback of that quantisation, as 
in the full theory, is that the unusual quantisation rule is highly ambiguous. One can take 
two extreme points of view: 1. If the LQC representation is chosen by nature, then  
some sort of unusual quantisation is enforced and one should find a selection criterion to 
find out which one is preferred. 2. The replacement of classically unbounded functions by bounded operators 
is a quantisation process nowhere else in physics to be found and therefore 
it is well motivated to look for representations different from the LQC representation 
without those unusual features. 

We therefore ask whether a quantum bounce is also possible 
in more standard representations, preferably derived from the full theory. 
For the present Fock incarnation, the cosmological truncation would 
rather lead to a regular representation of the Weyl algebra and a separable Hilbert space.
If we add the requirement of irreducibility then the representation is unitarily equivalent to 
the standard Schr\"odinger representation. Thus it would seem that the cosmological truncation 
of the present theory would yield the standard truncation of the Wheeler-DeWitt theory which 
does not display a bounce. In what follows, we show that this conclusion is not necessarily true.
Namely, one can use the non-polynomial nature of the gravitational Hamiltonian to argue that 
no "natural" operator ordering exists. Then one can use this ordering ambiguity to construct a 
Hamiltonian which produces a bounce, despite the representation being regular and the Hilbert space 
separable.

To show this, we use the simplest model discussed in the LQC literature which couples a homogenous 
real scalar field $\phi$ with conjugate momentum $\pi$ and vanishing potential 
to homogeneous isotropic gravity described by a scale factor $a\ge 0$ and conjugate 
momentum $p$. The only constraint left is the Hamiltonian constraint which for 
spatially flat metrics on a 3-torus of volume $L^3$ with suitable normalisations  
reads
\be \label{9.1}
C=\hbar\;L^3\; (-\frac{1}{\ell_P^2}\frac{p^2}{a}+\frac{\pi^2}{a^3})
\ee
with $a,\; L\; p, L^2\pi,\; L\phi$ dimension free (all spacetime 
coordinates carry length dimension). 
The non-trivial Poisson brackets are $L^3 \{p,a\}=\ell_P^2/\hbar,\; L^3\{\pi,\phi\}=1/\hbar$ and 
$\ell_P^2$ is the Planck area with a convenient numerical pre-factor of order unity

We construct the reduced phase space quantisation of this system and impose the gauge fixing condition
\be \label{9.2} 
\frac{\phi}{2\pi}+t=0
\ee
This condition is stable under evolution driven by $C$ provided that the dimension free 
lapse is $N=N_\ast=- a^3$. 
The solution of the constraint $C=0$ is $\pi^2=\pi_\ast^2=a^2 p^2/\ell_P^2$ and the reduced Hamiltonian 
acting on functions $F$ of $a,p$ is
\be \label{9.3}
\{H,F\}:=\{NC,F\}_{N=N_\ast,\pi=\pi_\ast,\phi=\phi_\ast}=\frac{\hbar L^3}{\ell_P^2} \{a^2 p^2,F\}        
%a^3 (-{p^2/a,F}+\pi_\ast^2/2 {1/a^3,F})=a^3(-1/a {p^2,F}+p^2/a^2 {a,F}-3a^2 p^2/a^4{a,F})
%=- a^3(1/a {p^2,F}+2p^2/a^2{a,F})=-2{a^2 p^2,F}
\ee
where $\phi_\ast=-2 t\pi_\ast$. Thus $H=\frac{\hbar L^3}{\ell_P^2} a^2 p^2$ and the reduced phase space is coordinatised by 
$a,p$.

In LQC one performs a canonical transformation on the reduced phase space which 
in the current notation would be given by $(a,p)\to (a^3,p/(3 a^2))$. In fact in LQC 
one works with a double cover of this phase space by switching to a homogenous and 
isotropic triad $e$ such that $e^2=a^2$ which has the advantage that $e$ can take 
real values while $a\ge 0$. 
Instead we keep the coordinates $(a,p)$ for the moment and consider the passage to 
the double cover later on. 
The Hamiltonian suffers from ordering ambiguities. Among the symmetric, positive operator 
valued orderings we consider 
two possibilities
\be \label{9.5}
H_1=\frac{\hbar L^3}{\ell_P^2}\; a \; p^2\; a,\;\; 
H_2=
\frac{\hbar L^3}{\ell_P^2} \;
\frac{a}{f(a)}\; p \; f(a)^2\;  p\; \frac{a}{f(a)}
\ee
where $f$ is a, for the moment arbitrary, function of the scale factor. 
Using the algebraic canonical commutation relations $[p,a]=i\ell_P^2/L^3$ 
valid in any representation we reorder $H_2$ to match $H_1$ and obtain the algebraic 
operator identity
%\be \label{9.6}
%H_2=H_1 + \frac{\ell_P^4}{b}
%\ee
%% ([1/b,q]+q 1/b) b^3 ([q,1/b]+1/b q)=(i l^2/b^2 +q /b ) b^3 (-i/b^2+1/b q)=(i l^2 b + q b^2)(-il^2 /b^2+1/b q)
%% = l^4/b+il^2 (q-q)+q b q
%\be \label{9.6}
%H_2=H_1+\ell_P^4 k k^{\prime\prime} h^2
%y k^2 y=([y,k]+ky)([k,y]+yk)=(i l^2 k'+ky)(-i l^2 k'+yk)
%=l^4 (k')^2+k y^2 k+il^2(k' yk-ky k')
%=l^4 (k')^2+k y^2 k+il^2(k' ([y,k]+ky)-([k,y]+yk) k')
%=l^4(k')^2+k y^2 k+il^2(k' (ik'+ky)-(-i k'+yk) k')
%=-l^4(k')^2+k y^2 k-il^2[y,k k'] 
%=l^4((k k')'-(k')^2)+k y^2 k
%=l^4 k k'' + k y^2 k
%\ee
%We now pick $k$ such that $k k{\prime \prime}$ is positive and such that
\be \label{9.6}
H_2=H_1+
%\frac{\hbar L^3}{\ell_P^2} (\frac{\ell^2_P/L^3})^3\; \frac{a^2 f^{\prime\prime}(a)}{f(a)}
\frac{\hbar \ell_P^2}{L^3}\; \frac{a^2 f^{\prime\prime}(a)}{f(a)}
\ee
% l^2=\ell_P^2/L^3
%p f^2 p=([p,f]+fp)([f,p]+pf)=(i l^2 f'+fp)(-i l^2 f'+pf)
%=l^4 (f')^2+f p^2 f+i l^2(f' pf- fpf') 
%=l^4 (f')^2+f p^2 f+i l^2(f' ([p,f]+fp) - ([f,p]+pf)f') 
%=l^4 (f')^2+f p^2 f+i l^2(f' (i l^2 f'+fp) - (-i l^2 f'+pf)f') 
%=-l^4 (f')^2+f p^2 f+i l^2(f'fp - p f f') 
%=-l^4 (f')^2+f p^2 f-i l^2([p, f f']) 
%=l^4 [(f f')')-(f')^2]+f p^2 f
%=l^4 f''+f p^2 f
The second term in (\ref{9.6}) is a normal ordering correction. It is produced by the 
same kind of ordering ambiguity as for the usual two ``natural´´, positive orderings of the 
harmonic oscillator $p^2+q^2, \; (q+ip)(q-ip)$ considered in the literature 
which in this case differ by a constant 
term (zero point energy). 

The function $f$ was left arbitrary so far. We now pick it in such a way that 
\be \label{9.7}
g^2:=\frac{a^2 f^{\prime\prime}(a)}{f(a)}
\ee
is a positive function which diverges to $+\infty$ as $a\to 0+$. Given $g$, (\ref{9.7})
can be considered as an ordinary differential equation of Riccati type for $u=f'/f$ 
\be \label{9.8}
u'+u^2=(\frac{g}{a})^2
\ee
Existence and uniqueness theorems grant the existence of maximal solutions for generic 
choices of $g$. 
We leave the discussion of generic choices of $g$ to the ambitious reader and confine 
ourselves to the remark that for 
the choice $g=\kappa/a$, with $\kappa$ a positive constant, this equation can be solved by
\be \label{9.9} 
f(a)=a\; e^{-\kappa a^{-1}}
\ee
% f'=(1+1/a) f, f´´=[-1/a^2+(1+1/a)1/a^2] f=1/a^3 f, f´´/f=1/a^4, a^2 f´´/f=1/a ^2
Note that all right handed derivatives exist and vanish at $a=0$. We can 
therefore extend $f$ to all of $\mathbb{R}$ by the symmetric choice $f(a):=f(-a)$ for 
$a<0$. Now all left handed derivatives exist and vanish as $a\to 0-$. Thus 
with this choice $f(a)$ is in fact smooth, positive and symmetric under reflection, it 
diverges linearly as $|a|\to\infty$.  
For completeness we note that the other linearly independent solution for $g=\kappa/a$ 
is given for $a>0$ by $a e^{\kappa a^{-1}}$, however it diverges together with all its derivatives at 
$a\to 0+$. 
% f=a e^{1/a}, f'=(1-1/a) e^{1/a}, f´´=1/a^2 f-(1-1/a)1/a^2 f =f/a^3, f´´/f=1/a^4

We may now argue as follows: We extend to the double cover by $a=|e|, p={\sf sgn}(e) q$ 
where $e,q$ are conjugate whence $e^2 q^2=a^2 p^2$. Since the Hamiltonian does not change 
under the substitution $a\to e, p\to q$ we abuse the notation and denote $e,q$ by $a,p$ again. 
We work in the standard Schr\"odinger representation ${\cal H}=L_2(\mathbb{R},da)$.
Since for non-linear Hamiltonians no ordering is sacred,
we adopt the ordering $H_2$ in the quantum theory. Since $H_2$ is symmetric and positive 
e.g. with dense invariant domain given by the states \cite{11}, it has self-adjoint extensions 
(e.g. the Friedrichs extension). Consider some initial 
state $\psi_0$ in the domain of $H_2$ and let $\psi_t :=e^{-i t H_2/\hbar}\; \psi_0$ its 
time evolution (solution of the Schr\"odinger equation with respect to the physical Hamiltonian).
Then 
\be \label{9.10}
<\psi_t,H_2\psi_t>= <\psi_0,H_2\psi_0>=
<\psi_t,H_1\psi_t>+ \kappa^2\frac{\hbar \ell_P^2}{L^3} <\psi_t,a^{-2} \psi_t>  
\ge \kappa^2\frac{\hbar \ell_P^2}{L^3} <\psi_t,a^{-2} \psi_t>  
\ee
where we used that both $H_1,H_2$ are in the domain \cite{11} and positivity of $H_1$. 
Thus we conclude that the expectation value of $a^{-2}$ remains bounded where the bound depends only on the 
choice of $\psi_0$ and the free parameter $\kappa$. This maybe interpreted as a quantum bounce.

In terms of the effective equations discussed in the LQC literature, if $\psi_0$ is 
a coherent state one might be able to show that the equations of motion for 
the expectation values of $a,p$ with respect 
to $\psi_t$ (Ehrenfest equations) can be derived from an effective Hamiltonian of the form 
\be \label{9.11}
H_2=\frac{\hbar L^3}{\ell_P^2} a p^2 a + \kappa^2\frac{\hbar \ell_P^2}{L^3} a^{-2} 
\ee
Since energy $E=H_2$ is conserved, in any solution of the Hamiltonian equations of motion 
we thus obtain similarly as in the quantum theory $a^{-2}\le E L^3/(\kappa^2\hbar\ell_P^2)$. 
Similarly, the effective Friedmann equations would give $\dot{a}=2 a^2 p$ and thus 
\be \label{9.12}
(\frac{\dot{a}}{a})^2=4 (ap)^2=4 \frac{\ell_P^2}{\hbar L^3}(E-\kappa^2\frac{\hbar \ell_P^2}{L^3} a^{-2})
\ee
In terms of the energy density of the scalar field given by 
$\rho=\hbar \pi^2/a^6=\hbar (a^2 p^2/\ell_P^2) 1/a^6=a^{-6}\; E/L^3$,
where the classical constraint, not corrected by the ordering ambiguity, was used, 
the energy density at the bounce is given by  
\be \label{9.12a}
\rho_{{\sf bounce}}=\frac{E}{L^3} (E L^3/(\kappa^2 \hbar\ell_P^2))^3=\rho_{{\sf Planck}}:=\frac{\hbar}{\ell_P^4}
\ee 
which can be fine tuned to the Planck density if we choose the free parameter $\kappa$ to be
\be \label{9.12b}
\kappa^6=(E L/\hbar)^4 (L/\ell_P)^2
% e:=EL/\hbar is dimension free, L/\ell_P=b is dimension free
% b^4 = e (e b^2 /(\kappa^2))^3=e^4 b^6 /\kappa^6
\ee
Clearly $\kappa$ needs to be chosen once and for all, thus one would pick the corresponding 
$E$ to be that energy that is measured at $t=0$ (today) and also adapt the coherent state 
to match that $E$. 
Note that we used a mechanism that is at the same time 1. entirely natural in quantum mechanics 
and 2. {\it the} distintive feature between classical and quantum mechanics, namely 
non-commutativity. This is possible while we use a regular representation and a separable Hilbert space.
However, apart from ordering ambiguities, which are always present for any Hamiltonian, there are no 
additional polymerisation quantisation ambiguities that result from replacements of e.g. $p$ by 
$\sin(\epsilon p)/\epsilon$ for some $\epsilon$ (which one can argue to be related to $\ell_P^2$) 
which are necessary due to the irregularity of the representation with respect to the 
1-parameter group of Weyl elements $s\mapsto e^{is p}$. In other words, also within the 
standard Schr\"odinger representation there exists a mechanism to obtain a quantum bounce 
which is very different from the mechanism at work in the LQC representation. It has the 
advantage that we do not need to replace the unbounded operator $p$ 
by the bounded $\sin(\epsilon p)/\epsilon$.

What does this imply with respect to the predictivity of the theory? 
One can criticise LQC that there are infinitely many polymerisations which in the limit 
$\epsilon \to 0$ reproduce the function $p^2$ and one can criticise the normal ordering argument 
by the fact that one can run the above reasoning with infinitely many functions $f$ such
that condition (\ref{9.7}) holds  (the correction terms are manifestly vanishing in the limit 
$\hbar\to 0$). One possibility is that input from the full theory could help to fix 
the polymerisation function or the normal ordering function using ideas from from 
renormalisation \cite{4}. Note, however, that the polymerisation ambiguity is 
{\it an additional ambiguity} that arises by the choice of an irregular representation:
also in LQC we could use further normal ordering functions besides the polymerisation 
ambiguity. In that sense, the ambiguities in irregular representations are larger than 
in regular representations.

\section{Conclusion} 
\label{s10}

We have carried out the same constraint quantisation steps as in LQG for a background independent 
Fock representation of geometry and scalar matter. It is important to stress that these 
steps are {\it at the same level of rigour as for LQG}. 
In fact, all the steps performed are rather standard, familiar to practitioners of QFT 
in Minkowski space. The Fock and LQG programme can be driven to similar levels of maturity. 

The main difference with LQG is that LQG also works for vacuum GR while the present 
quantisation relies on the presence of matter. As matter is an integral and experimentally 
verified part of GR, this reservation is not an obvious disadvantage. However, the matter 
that was used to construct the Fock representation should of course itself be experimentally 
verified. In the present paper we have used $D$ scalar fields. In the the real world with 
$D=3$ one such scalar field could be the Higgs field but two other scalar fields are required
for which experimental evidence is missing presently. Therefore in \cite{8} we assemble those 
missing scalar fields from other matter that is verified to be part of the standard model 
of elementary particle physics. 

Also in \cite{8} we consider the reduced phase space quantisation of the present theory 
which frees us from constructing the space of solutions of the constraints and the construction 
of the physical Hilbert structure on it but instead requires to quantise the reduced Hamiltonian 
as a self-adjoint operator. To construct an actual {\it operator} is even more difficult than for 
standard QFT in Minkowski space due to the non-polynomial structure of the reduced Hamiltonian. 
Thus we will content ourselves with the construction of a {\it quadratic form} using the techniques 
of section \ref{s8}. 

The fact that we used Fock space techniques raises the question whether one may apply 
perturbation theory to the reduced Hamiltonian, which is always a scalar density of 
weight one, in the sense that one can split it 
into a "free" and "interacting" part and then develop e.g. scattering theory in the standard 
perturbative way based on the Gell-Man-Low formula. This however appears to require to 
introduce background structure as otherwise there is no part of the Hamiltonian quadratic 
in the fields.

Finally the truncation to cosmology has revealed that in order to obtain a bounce {\it within 
the standard Sch\"odinger representation} 
we can make use of ordering ambiguities in defining the physical Hamiltonian which are 
familiar from the ordering ambiguities already for the harmonic oscillator. That such 
ordering ambiguities can have non-trivial physical imprints is well known and is best 
exemplified by the Casimir effect or the possibility that vacuum fluctuations in principle 
contribute to dark energy interpreted as a cosmological constant. These normal ordering 
ambiguities are always present in any representation and there is, to the best of our knowledge, no convincing 
argument in favour of one ordering or another beyond the subjective principle of "simplicity"
or "naturalness"". The advantage of regular representations is that there are no additional polymerisation 
parameter ambiguities and the Hilbert space is much smaller, allowing many standard constructions not 
possible in irregular representations. \\
\\
\\
{\bf Acknowledgements}\\
The present work was inspired by the rigorous canonical, background free and non-perturbative  
constraint quantisation of GR used in LQG. It was developed to a large extent by my friend 
and colleague Jurek Lewandowski together with various co-authors. Before Jurek's seminal 
contributions, canonical quantum GR was a mathematically poorly founded theory and without 
his work the field of LQG would not be at the present level of rigor and maturity. His 
input will be deeply missed. This article is dedicated to him.


\begin{thebibliography}{99}

\parskip -5pt

\bibitem{1} P. Hoehn, A.R.H. Smith, M.P.E. Lock. Trinity of relational quantum dynamics
Phys. Rev. {\bf D 104} (2021) 6, 066001. e-Print: 1912.00033 [quant-ph]\\
S. Carrozza, S. Eccles, P. A. Hoehn.
Edge modes as dynamical frames: charges from post-selection in generally covariant theories.
SciPost Phys. {\bf 17} (2024) 2, 048, SciPost Phys. {\bf 17} (2024) 048
e-Print: 2205.00913 [hep-th]\\
C. Goeller, P. A. Hoehn, J. Kirklin.
Diffeomorphism-invariant observables and dynamical frames in gravity: reconciling bulk locality with general covariance
e-Print: 2206.01193 [hep-th]

\bibitem{2}  C. Rovelli, ``Quantum Gravity'', Cambridge University
Press, Cambridge, 2004.\\
T. Thiemann, ``Modern Canonical Quantum General Relativity'', Cambridge
University Press, Cambridge, 2007\\
J. Pullin, R. Gambini, ``A first course in Loop Quantum Gravity'',
Oxford University Press, New York, 2011\\
C. Rovelli, F. Vidotto, ``Covariant Loop Quantum Gravity'', Cambridge
University Press, Cambridge, 2015\\
K. Giesel, T. Thiemann. Hamiltonian theory: dynamics.
In: Handbook of Quantum Gravity. C. Bambi, L. Modesto, I. Shapiro (eds.),
Springer Verlag, Berlin, 2023.

\bibitem{LQG1}  A. Ashtekar, ``New Variables for Classical and Quantum Gravity''
Phys. Rev. Lett. {\bf 57} (1986) 2244-2247\\
J. F. G. Barbero, ``A real polynomial formulation of general relativity in
terms of connections'', Phys. Rev. {\bf D49} (1994) 6935-6938

\bibitem{LQG2} A. Ashtekar, C.J. Isham, ``Representations of the Holonomy
Algebras of Gravity and Non-Abelean Gauge Theories'',
Class. Quantum Grav. {\bf 9} (1992) 1433, [hep-th/9202053]\\
Kinematical Hilbert spaces for Fermionic and Higgs quantum field theories.
Class. Quant. Grav. {\bf 15} (1998) 1487-1512; e-Print: gr-qc/9705021 [gr-qc]

\bibitem{LQG3} A. Ashtekar, J. Lewandowski, ``Representation
theory of analytic Holonomy $C^\star$ algebras'', in ``Knots and
Quantum Gravity'', J. Baez (ed.), Oxford University Press, Oxford 1994

\bibitem{LQG4} C. Fleischhack, ``Representations of the Weyl algebra in quantum geometry'',
Commun. Math. Phys. {\bf 285} (2009) 67-140, [math-ph/0407006]\\
J. Lewandowski, A. Okolow, H. Sahlmann, T. Thiemann,
``Uniqueness of diffeomorphism invariant states on holonomy-flux algebras''
Commun. Math. Phys. {\bf 267} (2006) 703-733, [gr-qc/0504147]

\bibitem{LQG5} A. Ashtekar, J. Lewandowski, ``Projective Techniques and
Functional Integration for Gauge Theories'', J.
Math. Phys. {\bf 36}, 2170 (1995), [gr-qc/9411046]

\bibitem{LQG6} A. Ashtekar, J. Lewandowski, D. Marolf, J. Mourao, T.Thiemann.
Quantization of diffeomorphism invariant theories of connections with local degrees of freedom
J. Math. Phys. {\bf 36} (1995) 6456-6493. e-Print: gr-qc/9504018 [gr-qc]

\bibitem{LQG7} T. Thiemann, ``Anomaly-free Formulation of non-perturbative,
four-dimensional Lorentzian Quantum Gravity'', Physics Letters {\bf B380}
(1996) 257-264, [gr-qc/9606088]

\bibitem{LQG8} Ashtekar, J. Lewandowski, D. Marolf, J. Mourao, T.Thiemann.
Coherent state transforms for spaces of connections.
J. Funct. Anal. {\bf 135} (1996) 519-551. e-Print: gr-qc/9412014 [gr-qc]\\
T. Thiemann, ``Complexifier coherent states for canonical
quantum general relativity'',
Class. Quant. Grav. {\bf 23} (2006) 2063-2118,
[gr-qc/0206037]

\bibitem{LQG9} A Length operator for canonical quantum gravity.
J. Math. Phys. {\bf 39} (1998) 3372-3392. e-Print: gr-qc/9606092 [gr-qc]\\
C. Rovelli and L. Smolin.
Discreteness of volume and area in quantum gravity.
Nucl. Phys. {\bf B442} (1995), 593-622; Erratum: Nucl. Phys.
{\bf B456} (1995) 753, [gr-qc/9411005]\\
A. Ashtekar and J. Lewandowski.
Quantum theory of geometry I: Area Operators.
Class. Quant. Grav. {\bf 14} (1997) A55-A82, [gr-qc/9602046]\\
A. Ashtekar and J. Lewandowski.
Quantum theory of geometry II:
Volume operators. Adv. Theo. Math. Phys. {\bf 1} (1997) 388-429,
[gr-qc/9711031]

\bibitem{3} K. Giesel, T. Thiemann, ``Scalar Material Reference Systems and Loop Quantum
Gravity'', Class. Quant. Grav. {\bf 32} (2015) 135015,
[arXiv:1206.3807]

\bibitem{4} T. Thiemann. 
Canonical Quantum Gravity, Constructive QFT, and Renormalisation.
Front. in Phys. {\bf 8} (2020) 548232, Front.in Phys. {\bf 0} (2020) 457.
e-Print: 2003.13622 [gr-qc]

\bibitem{5} S. Fulling. Aspects of Quantum Field Theory in
Curved Spacetime.
London Math. Society Student Texts, vol. 17, 1989.

\bibitem{6} T. Thiemann.
Observations on representations of the spatial diffeomorphism group and algebra in all dimensions
Phys. Rev.D {\bf 110} (2024) 12, 124022. e-Print: 2405.01201 [gr-qc]

\bibitem{7} T. Thiemann. Non-perturbative quantum gravity in Fock representations
Phys. Rev. D {\bf 110} (2024) 12, 124023. e-Print: 2405.01212 [gr-qc]

\bibitem{8} T. Thiemann. Non-perturbative, background independent canonical quantum gravity
in Fock representations. Companion paper.

%\bibitem{8a} T. Thiemann, O. Winkler.
%Gauge Field Theory Coherent States (GCS) : IV. Infinite Tensor Product and Thermodynamical Limit.
%Class. Quant. Grav. {\bf 18} (2001) 4997-5054. arXiv: hep-th/0005235.

\bibitem{9} A. Ashtekar, P. Singh,
``Loop Quantum Cosmology: A Status Report'',
Class. Quant. Grav. {\bf 28} (2011) 213001, [arXiv:1108.0893]\\
I. Agullo, P. Singh, ``Loop Quantum Cosmology'', [arXiv:1612.01236]

\bibitem{10} M. Bojowald. Quantum cosmology: a review. Rept. Prog. Phys. {\bf 78} (2015) 023901.
e-Print: 1501.04899 [gr-qc]

\bibitem{11}  T. Thiemann.
Properties of a smooth, dense, invariant domain for singular potential 
Schroedinger operators. e-Print: 2305.06718 [quant-ph]\\
J. Neuser, T. Thiemann.
Smooth, invariant orthonormal basis for singular potential Schroedinger 
operators. e-Print: 2308.07059 [quant-ph]



\end{thebibliography}
\end{document}